\documentclass[10pt,journal,compsoc]{IEEEtran}
\ifCLASSOPTIONcompsoc
  \usepackage[nocompress]{cite}
\else
  \usepackage{cite}
\fi
\ifCLASSINFOpdf
\else
\fi

\usepackage{hyperref}
\usepackage{graphicx}
\usepackage{xcolor}
\usepackage{amsmath}
\usepackage{multirow}
\usepackage{lipsum}
\usepackage{diagbox}
\usepackage{tabularx}

\newcommand{\revision}[1]{\textcolor{black}{#1}}

\begin{document}
%
\title{On Rotation Gains Within and Beyond Perceptual Limitations for Seated VR}
%
%
%
%

\author{
Chen~Wang,~\IEEEmembership{Student Member,~IEEE,}
Song-Hai~Zhang,~\IEEEmembership{Senior~Member,~IEEE,}
Yizhuo~Zhang,
Stefanie Zollmann,~\IEEEmembership{Member,~IEEE,}
and~Shi-Min~Hu,~\IEEEmembership{Senior~Member,~IEEE}
\IEEEcompsocitemizethanks{
\IEEEcompsocthanksitem Chen Wang is with BNRist, Department of Computer Science and Technology, Tsinghua University. Email: \href{mailto:cw.chenwang@outlook.com}{cw.chenwang@outlook.com}
\IEEEcompsocthanksitem Song-Hai Zhang is with BNRist, the Department of Computer Science and Technology, Tsinghua University. Email: \href{mailto:shz@tsinghua.edu.cn}{shz@tsinghua.edu.cn}
\IEEEcompsocthanksitem Yizhuo Zhang is with BNRist, Department of Computer Science and Technology, Tsinghua University. Email: \href{mailto:yizhuo-z18@mails.tsinghua.edu.cn}{yizhuo-z18@mails.tsinghua.edu.cn} 
\IEEEcompsocthanksitem Stefanie Zollmann is with the Department of Computer Science, University of Otago. Email: \href{mailto:stefanie.zollmann@otago.ac.nz}{stefanie.zollmann@otago.ac.nz}
\IEEEcompsocthanksitem Shi-Min Hu is with BNRist, the Department of Computer Science and Technology, Tsinghua University. Email: \href{mailto:shimin@tsinghua.edu.cn}{shimin@tsinghua.edu.cn}  
}
}

%
%

\markboth{Journal of \LaTeX\ Class Files,~Vol.~14, No.~8, August~2015}%
{Shell \MakeLowercase{\textit{et al.}}: Bare Demo of IEEEtran.cls for Computer Society Journals}
%



\IEEEtitleabstractindextext{%
\begin{abstract}
Head tracking in head-mounted displays (HMDs) enables users to explore a 360-degree virtual scene with free head movements. However, for seated use of HMDs such as users sitting on a chair or a couch, physically turning around 360-degree is not possible. Redirection techniques decouple tracked physical motion and virtual motion, allowing users to explore virtual environments with more flexibility. In seated situations with only head movements available, the difference of stimulus might cause the detection thresholds of rotation gains to differ from that of redirected walking. Therefore we present an experiment with a two-alternative forced-choice (2AFC) design to compare the thresholds for seated and standing situations. Results indicate that users are unable to discriminate rotation gains between 0.89 and 1.28, a smaller range compared to the standing condition. We further treated head amplification as an interaction technique and found that a gain of 2.5, though not a hard threshold, was near the largest gain that users consider applicable. Overall, our work aims to better understand human perception of rotation gains in seated VR and the results provide guidance for future design choices of its applications.
\end{abstract}

}

\maketitle

\IEEEdisplaynontitleabstractindextext

%
\IEEEpeerreviewmaketitle


\section{Introduction}
\IEEEPARstart{V}{}irtual reality (VR) technology has developed significantly over recent years, providing immersive virtual environments (VEs) for various applications including games, training, and rehabilitation \cite{VRSurvey}. Consumer-level head-mounted displays (HMDs) such as HTC Vive, Valve Index, Oculus Quest, equipped with accurate head and body tracking techniques, enable users to explore virtual scenes more freely than ever before. However, realistic physical movements are still a fundamental concern in VR, which has been proven to be essential for understanding and interacting with VEs~\cite{usoh1999walking, ragan2016amplified}. The most common way is a one-to-one mapping that 
alters the virtual position in the exactly same manner as real ones. Although providing a high level of consistency and naturalness\cite{bowman2012questioning}, it is not the desired way in many VR situations. Recovering movements in fidelity might not be possible due to the absence of large physical space. More importantly, VR is increasingly used for home and public entertainment purposes covering a wider range of users, in which scenarios users are mostly seated and prefer more convenient and relaxed interactions while excess body movements might be tiring and uncomfortable. Existing surveys also demonstrate that sitting provides more comfort and causes fewer safety concerns in leaning-based interfaces~\cite{zielasko2021sit}.

This motivates research into techniques that facilitate users to explore and navigate the VE when being seated. Classical methods include steering with directions given by gaze, torso or upper body~\cite{zielasko2020take}, walk-in-space techniques~\cite{zielasko2016evaluation} or manipulation with hand-based input devices, but still requires a swivel chair for 360-degree viewing. Another solution is to utilize modified interaction techniques for better travel and view control, specifically redirection techniques. Based primarily on the visual dominance effect\cite{burns2005hand}, these works manipulate the mapping between physical movements and virtual ones. For instance, \textit{redirected walking}\cite{razzaque2005redirected, steinicke2008analyses} allows users to explore virtual worlds that are much larger than the actual physical space by slightly bending the walking path without being noticed. In terms of head movements, users can view a 360-degree virtual scene with much less head motion with amplified head rotation~\cite{jay2003amplifying, le2013evaluating, ragan2016amplified, sargunam2017guided}, which also provides higher-fidelity control than joystick or mouse techniques. However, unsuitable amplification factors have been found to damper task performance and introduce sickness~\cite{ragan2016amplified, sargunam2017guided}. 

Obviously, head rotations cannot be changed in an arbitrary way. Previous works have estimated the detection thresholds (the range within which users cannot detect the manipulation of a certain redirection technique) of several types of gains for redirected walking~\cite{steinicke2009estimation, bruder2012redirecting}. During their experiment for estimating rotation gains, either in-place or during walking, users can turn their bodies freely when moving their heads. \revision{However, in seated situations such as non-turning chairs and sofas, although upper body movement is available, people may prefer only to rotate their heads as it potentially offers a more convenient and comfortable experience.} We thus revisit the problem of estimating the range of imperceptible rotation gains for seated VR, first studied by~\cite{jaekl2005perceiving}, with modern hardware and controlled upper body motion. In addition, previous work found that rotation gains within the upper detection thresholds are generally less than 1.5 (e.g. results in~\cite{steinicke2009estimation, williams2019estimation}) and far from sufficient for 360-degree exploration for stationary use. Also, while overt manipulation has already been studied in head rotation, researchers have not well-explored users' subjective preferences to those large gains and how strong a user might be manipulated. This necessitates further research into the understanding of rotation gains beyond perceptual limitations.

Therefore, our research focuses on answering the following questions:
\begin{itemize}
    \item \textbf{Q1}: What are the detection thresholds of rotation gains for seated VR restricted to head motion only?
    \item \textbf{Q2}: To what extent can users accept gains larger than the detection threshold when seated?
\end{itemize}

In this paper, we address the above questions with two user studies. The first study is a psychophysical experiment utilizing a two-alternative forced-choice methodology and estimating detection thresholds of rotation gains in seated VR with psychometric functions. \revision{We confirmed that compared to the standing condition, people are even more sensitive to manipulations while seated. To bridge the gap between imperceptible gains and real-world applications, we further investigate whether gains beyond the threshold can be tolerated.} Therefore, in the second study, we use a set of applicability items including naturalness, and sickness and asked participants to give their subjective ratings that were later used for comparison between different levels of gains. We revealed that most participants thought gains up to 2.5 were practical for usage. Our findings provide valuable insights and open up further research directions for seated VR applications.
\section{Related Work}
Our work builds on previous research about natural and semi-natural navigation in VR and human perception of them. In this section, we provide an overview of current advances on redirection techniques, threshold detection, as well as how our research relates to them.

\subsection{Redirection Techniques}
Redirection techniques have to date been investigated in many situations such as redirected walking ~\cite{razzaque2005redirected, steinicke2008analyses, stebbins2019redirecting, razzaque2002redirected}, hand remapping and redirected touching~\cite{zenner2019estimating, benda2020determining, esmaeili2020detection, kohli2012redirected, matthews2019remapped}, jumping~\cite{hayashi2019redirected, liu2021redirected} and head rotations ~\cite{sargunam2017guided, stebbins2019redirecting}. The main goal of these techniques is to overcome the problems of limited space or physical incapabilities~\cite{WANG2021iv}. If only one-to-one mapping is employed, users' walking will be constrained by the room space and arm reach by their arm length.

Redirected walking (RDW), proposed by Razzaque et al.~\cite{razzaque2005redirected}, is a common locomotion technique for natural traveling in VEs without joystick controls. Rotations and translations are applied imperceptibly so that users adjust their movements to walk along a curved physical path as if to move on a virtual straight line. The ratio of altered position and tracked physical position is defined as gain~\cite{steinicke2009estimation} or control/display ratio~\cite{frees2005precise}. Langbehn et al.~\cite{langbehn2017application} suggested that RDW also enabled users to walk on already curved paths and introduced the concept of bending gain. Researchers later investigated how to make RDW more applicable in various ways. Razzaque et al.~\cite{razzaque2002redirected} examined redirection in CAVE-like environments that are not equipped with 360-degree screens. Their method allowed users to virtually move around without facing the missing back wall, which was achieved by gradually rotating the virtual scene towards the direction of the front screen. Seven League Boots~\cite{interrante2007seven} instead amplified movements considering both user gaze direction and travel direction to increase walking distance. Studies also devised RDW algorithms for gymnasium-sized spaces~\cite{hodgson2008redirected}. To encourage users to steer towards predefined locations, guidance fields~\cite{tanaka2016guidance} could be used. Additionally, the reorientation process could be less obvious to users when distractors were placed in the VE~\cite{peck2011evaluation}. 

However, physical walking is not always possible, in many VR applications the user is seated or standing with only head and upper body motion available. Recently, novel techniques were proposed to tackle this issue, i.e. manipulation of lateral camera motion in response to the corresponding head motion in VR~\cite{serrano2020imperceptible}, or translating the motion of in-place pedaling to virtual walking~\cite{freiwald2020walking}. 
Other researched approaches include hand and head redirection techniques~\cite{poupyrev1996go, frees2005precise, li2015evaluation, kohli2012redirected, cheng2017sparse, han2018evaluating}. Highly related to our work is redirected head rotation, especially for seated VR. This can be done through head scrolling as introduced in ~\cite{zielasko2016evaluation} where virtual scenes would rotate towards the center view when the head rotation angle surpassed a predetermined threshold. Norouzi et al.~\cite{norouzi2019augmented} further extended it by using eye movements to control the virtual view angle.
Another line of work is redirected head movements with amplified head rotations that introduce less sickness than scrolling ~\cite{langbehn2019turn}.
Head rotations are amplified using an amplification factor (or rotation gains) to allow viewing for a large range with relatively small physical turns. Le Ngoc et al.~\cite{le2013evaluating} explored rotation amplification for flight simulation and concluded that no significant extra workload or simulator sickness would be induced. Jay et al.~\cite{jay2003amplifying} demonstrated that amplified head rotation even improved the performance of visual search tasks. Freitag et al.~\cite{freitag2016examining} detected no significant negative effect of rotation gains on simulator sickness and presence in CAVE systems but found evidence of reduced spatial knowledge. Further experiments that compared four levels of amplification factors in both CAVE and HMD systems were conducted by Ragan et al.~\cite{ragan2016amplified}. Although the performance on search tasks was not affected by amplification, they found obvious sickness problems and worse spatial orientation with large rotation gains. Instead of using a constant mapping, rotation gains can be altered dynamically without being noticed by users~\cite{zhang2013human}. Langbehn et al.~\cite{langbehn2019turn} showed that dynamic gains caused less sickness and had higher usability. Zhang et al.~\cite{zhang2021velocity} designed a velocity-guided amplification function that outperformed linear mapping on visual searching and counting tasks. Sargunam et al.~\cite{sargunam2017guided} combined dynamically amplified and guided head rotations that allowed a full 360-degree virtual range exploration with comfortable physical rotation, whose rotation adjustments resembled those used in washout filters for motion simulation~\cite{huang2010human, wang2004predictive}.


Although previous works enhanced user experience or facilitated performance, there is generally a lack of understanding of how much rotation gain is imperceptible or applicable for seated VR. Rotation gains for RDW provide a reasonable reference, but considering the flexibility difference between the two, a comparison of them will certainly provide precious information for devising amplification techniques for seated VR applications.

\subsection{Detection Thresholds}
Redirection techniques cannot be applied without consideration of magnitude, therefore studies have applied psychophysical methods to determine detection thresholds. The methodology widely used recently is to present participants a certain level of stimuli in each trial and force them to choose from two options (i.e., Was the virtual movement \textit{larger} or \textit{smaller} than the physical movement?). The proportion of correct answers for each level is then calculated and fitted to a psychometric function. Based on this, detection thresholds can then be estimated. This procedure is typically called 2AFC and employed in a lot of previous literature that also focuses on threshold estimation of redirection techniques ~\cite{steinicke2009estimation, benda2020determining, esmaeili2020detection, zenner2019estimating, ogawa2020effect}. Grechkin et al.~\cite{grechkin2016revisiting} proposed to name it pseudo-2AFC since only one stimulus is presented in each trial but in our paper we still follow the terminology 2AFC.



Estimating rotation gains in redirected walking has been well studied. Steinicke et al.~\cite{steinicke2009estimation} found a detection threshold for rotation gains for redirected walking between 0.64 and 1.24 (lower and upper bound) by 2AFC. Bruder et al.~\cite{bruder2012redirecting} reported similar gains at 0.68 and 1.26. The same authors~\cite{bruder2009impact} detected no significant differences in 13 participants between two gender groups, 0.69 and 1.19 for men, 0.66 to 1.26 for women. Other researchers estimated rotation gains with varying experimental conditions. Paludan et al.~\cite{paludan2016disguising} studied threshold gains in the presence of 4 objects and 16 objects and found values at 0.81 - 1.19 and 0.82 - 1.20 respectively. When visual cues were presented, thresholds for rotation gain with or without audio were revealed are similar~\cite{nilsson2016estimation}. Serafin et al.~\cite{serafin2013estimation} estimated detection thresholds to be at 0.28 to 1.2 with only auditory stimuli and Freitag et al.~\cite{freitag2016examining} derived thresholds from 0.85 to 1.18 for CAVE-like environments. Williams and Peck~\cite{williams2019estimation} investigated rotation gains considering FOV, gender and distractors. \revision{Recently, thresholds under controlled rotational speeds were also tested and participants were less sensitive to rotation gains with decreased rotational motion~\cite{brument2021studying}.}

In the context of rotation gains for seated conditions, Jaekl et al.~\cite{jaekl2005perceiving} first conducted experiments that asked seated participants to adjust rotation gains by a step of 0.05 until the VE was stable. They found detection thresholds for the yaw axis of 0.88 and 1.33 in a 10 participants experiment.
Jerald et al.~\cite{jerald2008sensitivity} reported 2D scenes presented by a screen can be rotated up to 11.2\% with the direction of head rotation and 5.2\% against the direction of head rotation in a seated setting. Bruder et al.~\cite{bruder2012redirecting} also conducted experiments for seated conditions and reported gains of 0.77 to 1.26. However, their experiments required participants to sit on a rotatable wheelchair and use joysticks for initiating rotations and they might get proprioceptive cues from their hands in this way. Our work resembles the work of Jaekl et al.~\cite{jaekl2005perceiving} as we also measure seated rotation gains but differs in the aspect that our experiment requires users to rotate their heads without upper body movement, which is not explicitly controlled in their work~\cite{jaekl2005perceiving}.

\subsection{Overt Manipulation}
According to Suma et al.~\cite{suma2012taxonomy}, redirection techniques can be classified as overt and subtle in terms of noticeability to users. While subtle methods have the benefit of being less noticeable, they have limited potential to improve the overall user experience (i.e. reduce the physical movement to a considerable amount). This can be achieved by overt manipulation such as teleportation, resetting~\cite{williams2007exploring}, or use of perceptible gains. Rietzler et al.~\cite{rietzler2018rethinking} confirmed that users could accept curvature gains far beyond perceptual thresholds in RDW. Simeone et al.~\cite{simeone2020space} designed another overt walking technique in the context of room-scale VR, which was faster and preferable than compared techniques. Concurrently, Telewalk~\cite{rietzler2020telewalk} combined observable translation and curvature gains to allow endless virtual walking along a real-world path with a pre-defined radius. However, these techniques might break the sense of presence and introduce extra motion sickness. Furthermore, Schmitz et al.~\cite{schmitz2018you} studied the amount of rotation gain that reduces self-reported presence through a search and collect task. They showed that rotation gains above detection thresholds still allow providing immersive experiences.

For seated VR, rotation gains within the detection threshold are not sufficient for 360-degree virtual viewing, since head rotation of at most 90 degrees is available~\cite{langbehn2019turn}. Many existing head amplification techniques can be seen as overt, where they either use a large rotation gain constantly~\cite{langbehn2019turn, ragan2016amplified, kopper2011towards} or in part of the remapping curve~\cite{sargunam2017guided, langbehn2019turn}. As reported by Ragan et al.~\cite{ragan2016amplified}, noticeable problems were identified with an amplification factor of 4. Inspired by these studies, we aim to determine the threshold that users are aware of but still find acceptable to use.

\section{Experiment 1: Rotation Gain Thresholds}
The goal of the first experiment is to determine the detection thresholds of rotation gains when users are seated restricted to head movements only. Before starting the experiments and analyzing the rotation gains, we were interested in estimating the angles that users feel comfortable in such a scenario. 

\begin{figure}[htbp]
    \centering
    \includegraphics[width=0.75\linewidth]{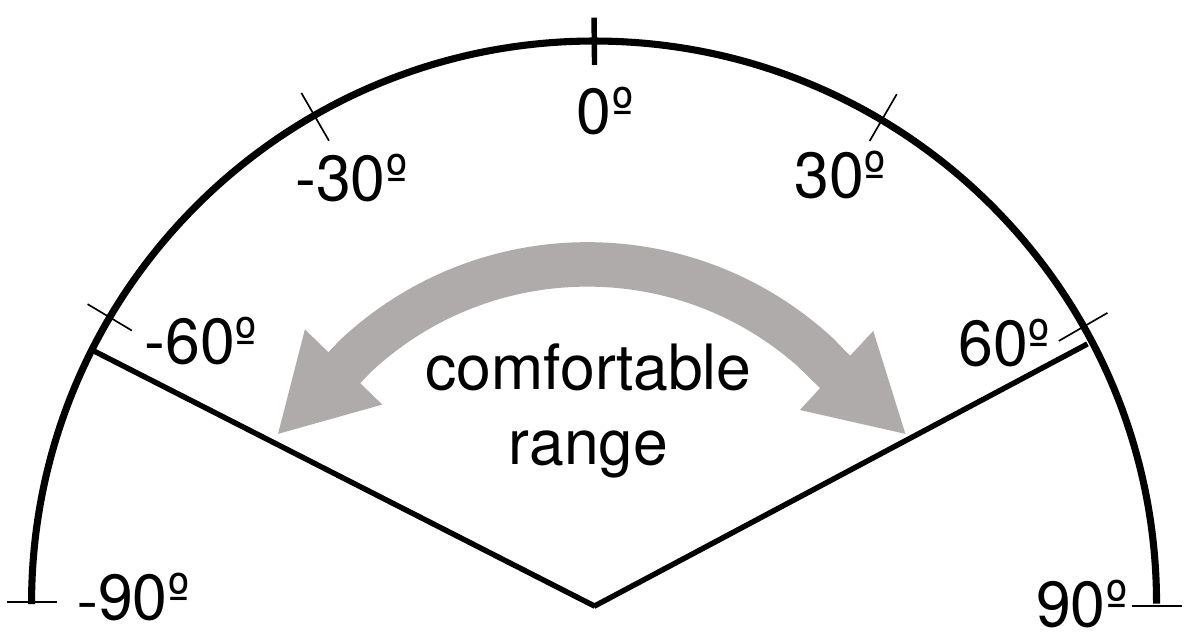}
    \caption{A top-down diagram showing the maximum comfortable rotation angle, which is less than 90 degrees in both directions. }
    \label{fig:maximum-angle}
\end{figure}

When users are stationary seated, the largest physical angle that they can perform without twisting their body both on the left and right side is less than 90$^{\circ}$ \cite{langbehn2019turn}. However, we were interested to determine the angle that they feel comfortable with, which we anticipate to be smaller. We define this angle as the \textit{maximum comfortable rotation angle}. In order to quantify this angle, we conducted a pilot study. We asked 10 participants (6 male and 4 female, mean age 21.7) to rotate their heads to the maximum comfortable rotation angle using a VR headset. We asked them to perform this task three times both in the left and right direction and captured the angle from the headset sensors. We then calculated the averages for the maximum comfortable rotation angle for left 62.5$^{\circ}$ (SD = 9.6) left and for right 61.6$^{\circ}$ (SD = 10.8) (Fig. \ref{fig:maximum-angle}). Values ranged from 43$^{\circ}$ to 76$^{\circ}$, with 6 participants' angles were in the range of 60$^{\circ}$ to 70$^{\circ}$. This indicates that many users prefer smaller head movements than what they are physically capable of.
Note that measures including wearing masks, cleaning the equipment before and after experiments and keeping a safe social distance were taken to prevent the spread of COVID-19 in all the experiments in this paper.

\subsection{Experiment Design}
The main goal of this experiment is to estimate the detection thresholds of rotation gains in a seated scenario. For the experiment, we deployed a 2AFC task during which a random gain was applied constantly in each trial and the yaw axis of the head rotation in the VR HMD was amplified accordingly. The forced-choice aims to avoid bias because when subjects do not know the answer, they will have a probability of 50\% to be correct\cite{steinicke2009estimation}. The rotation angle in the real world was randomized between $60^{\circ}$ and $70^{\circ}$ with an average rotation angle of $65^{\circ}$ to confine the movement close to most users' maximum comfortable rotation angle. We did not include angles less than 60 since participants were not able to notice if angles were lower within 10 degrees. The rotation also would be not enough and required more turn around.

The experiment contains two blocks: \textit{seated} and \textit{standing} which only differ in the posture during the rotation and the order of blocks were counterbalanced between participants. In the seated block, participants only rotated their heads, while in the standing block they could move their bodies freely. Rotation gains ranged from 0.5 to 1.5 with steps of 0.1 were included. Each gain was repeated six times and participants had to finish a total of 4 random practice trials + 11 gains $\times$ 6 repeats = 70 trials in both blocks. The trial ordering was randomized for each participant. Answers and response times (from the time the question was shown to the time they made the selection) of the participants were logged for analysis.

For each trial, participants rotated their heads in the direction of a virtual arrow that appeared at the center of their vision. They rotated until hitting a red cylinder, indicating that they should turn their heads backward till they noticed another cylinder at the center of their vision. Then, they needed to rotate to the neutral position and the trial was finished. The experiment design allowed participants to experience enough virtual rotation and avoided head strain since the maximum angle was 70$^{\circ}$ and the final head rotation angle was close to zero. 

Once a trial ended, the HMD would fade to light navy and participants were prompted with the question ``Was the virtual movement smaller or greater than the physical movement?'' (smaller, greater) and had to submit their answer with a hand controller. After making the choice, they needed to select the ``Next'' option in the scene to proceed. The virtual environment in the next trial would be rotated at a random angle to avoid familiarization with the orientation. If a break was needed, ``Next'' should remain unselected before the experiment resumed. To avoid possible distractions or hints, the arrow and cylinders would disappear after participants started to rotate in the right direction. Participants were encouraged to rotate with a normal speed without overshooting the cylinder position. The accuracy of their response was not given in any trial.

\begin{figure}[htbp]
    \centering
    \includegraphics[width=0.95\linewidth]{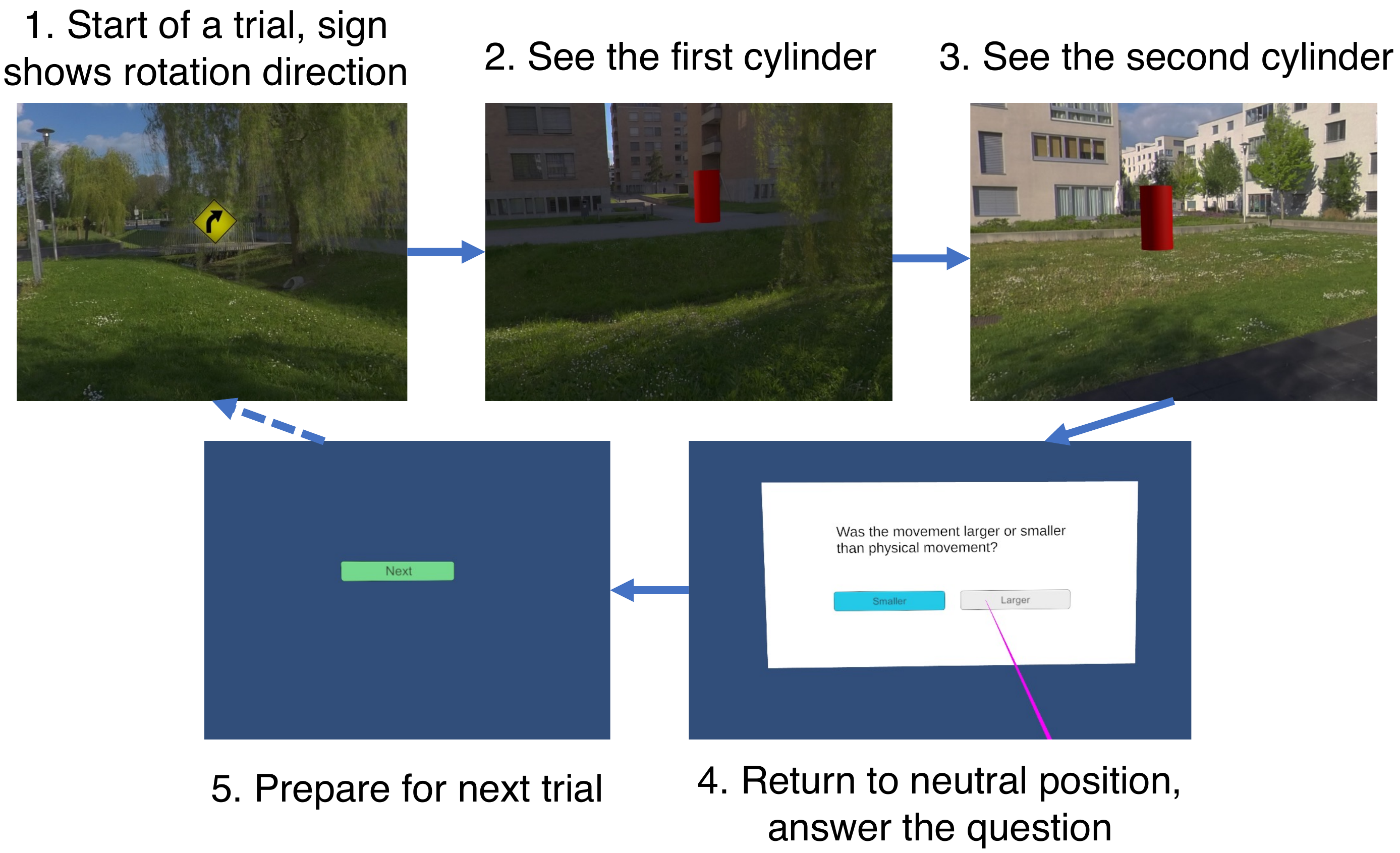}
    \caption{First-person view of a trial in the experiment.}
    \label{fig:exp1-view}
\end{figure}

The virtual scene consisted of a city block with blue sky, grasses, trees, roads and buildings with different colors. The participant was positioned at a lane and in the center of the buildings. The environment had rich textures and realistic shadows that provided sufficient optical flow when users rotated their heads (See Fig. \ref{fig:exp1-view} for the virtual scene and an exemplar process of a trial).

\subsection{Apparatus}
\begin{figure}[htbp]
    \centering
    \includegraphics[width=1.0\linewidth]{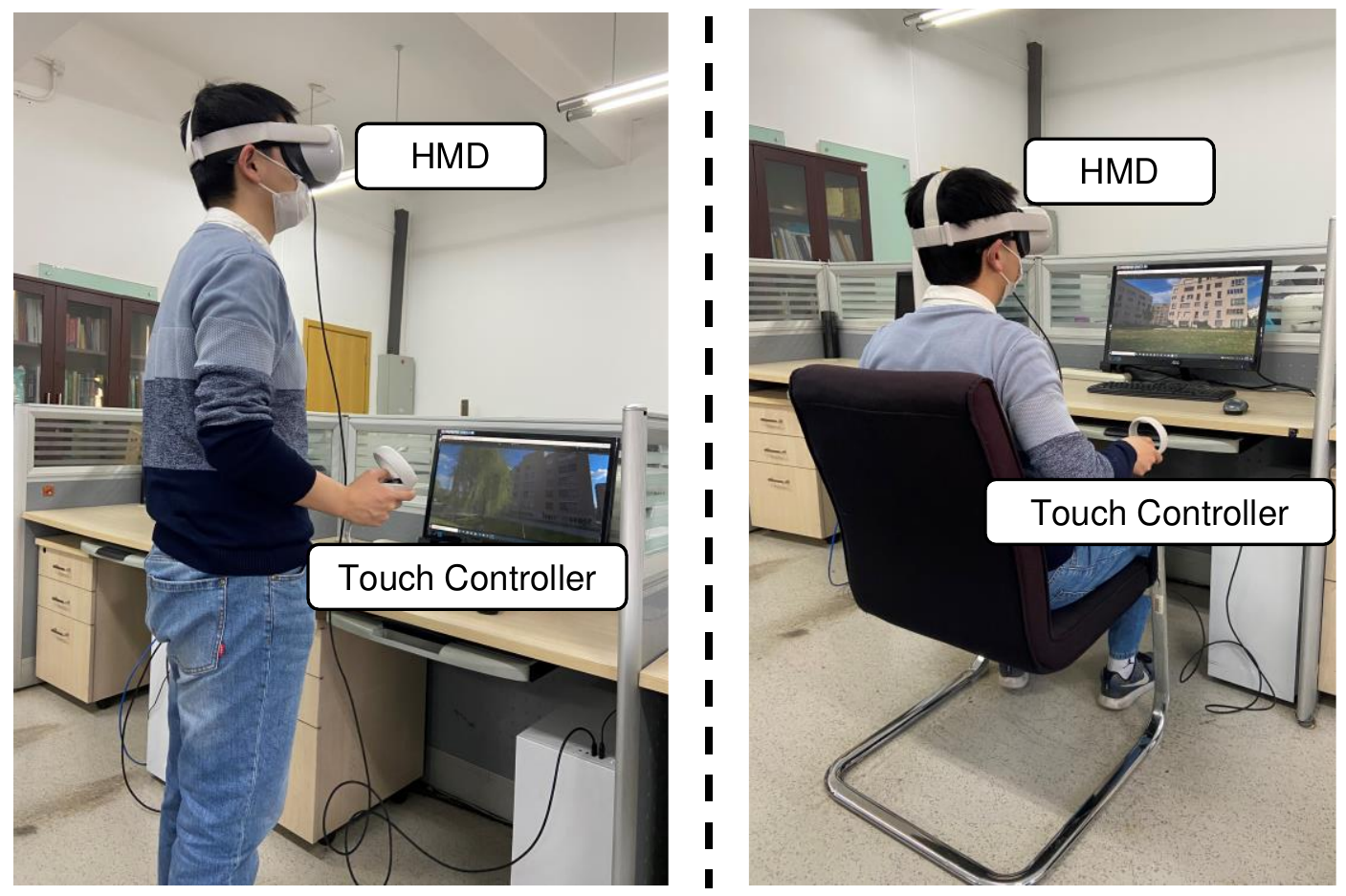}
    \caption{Setup for the first experiment in the standing block (left) and seated block (right).}
    \label{fig:exp1-setup}
\end{figure}

We used an Oculus Quest 2 headset with the default 6DOF position and orientation tracking system, as well as the right-hand controller since all participants chose to use their right hands to answer the questions. The interpupillary distance (IPD) of the HMD (3 levels for Oculus Quest 2) was calibrated for each participant to avoid that the display was blurry. The system has around 110$^{\circ}$ diagonal FOV, a 90Hz refresh rate and a 1832 $\times$ 1920 resolution per eye. The software was developed using Unity3D 2020.3.12f1c1 64bit (with Oculus Integration plugin). During the experiment, the headset was connected to a desktop computer with a cable which was long enough to not hamper movement, so the experimenter was able to monitor the virtual scene mirrored on the computer display. Ambient background sounds were played to avoid the orientation cues introduced by real-world sounds. For the seated block, participants sat on a non-rotating chair with little movements of the upper body. For the standing block, participants stood in-place and rotated their heads and upper bodies freely. The setup is shown in Fig. \ref{fig:exp1-setup}.

\subsection{Participants}
20 subjects, age 19-25, including 10 male ($M=21.6$, $SD=2.17$) and 10 female ($M=21.3$, $SD=2.26$) successfully accomplished the study,  3 additional participants finished the experiment but were excluded from data analysis (See Sec. \ref{subsec:exp1-res}). Most of them were students or members from the campus and had diverse backgrounds. All participants had a normal or corrected-to-normal vision (13 wear glasses or contact lenses) and were physically and mentally healthy for the experiment. 4 participants had multiple VR experiences before, 8 had experienced some VR before, and 8 had never experienced VR. 

\subsection{Procedure}

Upon their arrival at the lab, participants filled in a consent form and a demographic questionnaire (age, gender, amount of prior experience with VR). Each of them received \$20 for compensation. Then the experimenter explained the goal and task of the experiment. Participants either started with the standing or seated block after the experimenter helped them adjust and put on the HMD. At the beginning of a block, four practice trials were included for participants to get familiar with the process during which they were allowed to ask questions. Afterwards, they began to repeat the remaining formal trials with head rotation and 2AFC selection. Participants were allowed to take a short break after each trial that lasted no longer than three minutes. After completing all trials of the first block, they would have a 10-minute compulsory break and then proceed to the second block. A Kennedy-Lane Simulator Sickness Questionnaire (SSQ)\cite{kennedy1993simulator} was completed before and after each block. The entire procedure took about 90-100 minutes including explanation, trials, breaks and questionnaires. 

\subsection{Results}
\label{subsec:exp1-res}
\subsubsection{Detection Thresholds}
\begin{figure}[htbp]
    \centering
    \includegraphics[width=1.0\linewidth]{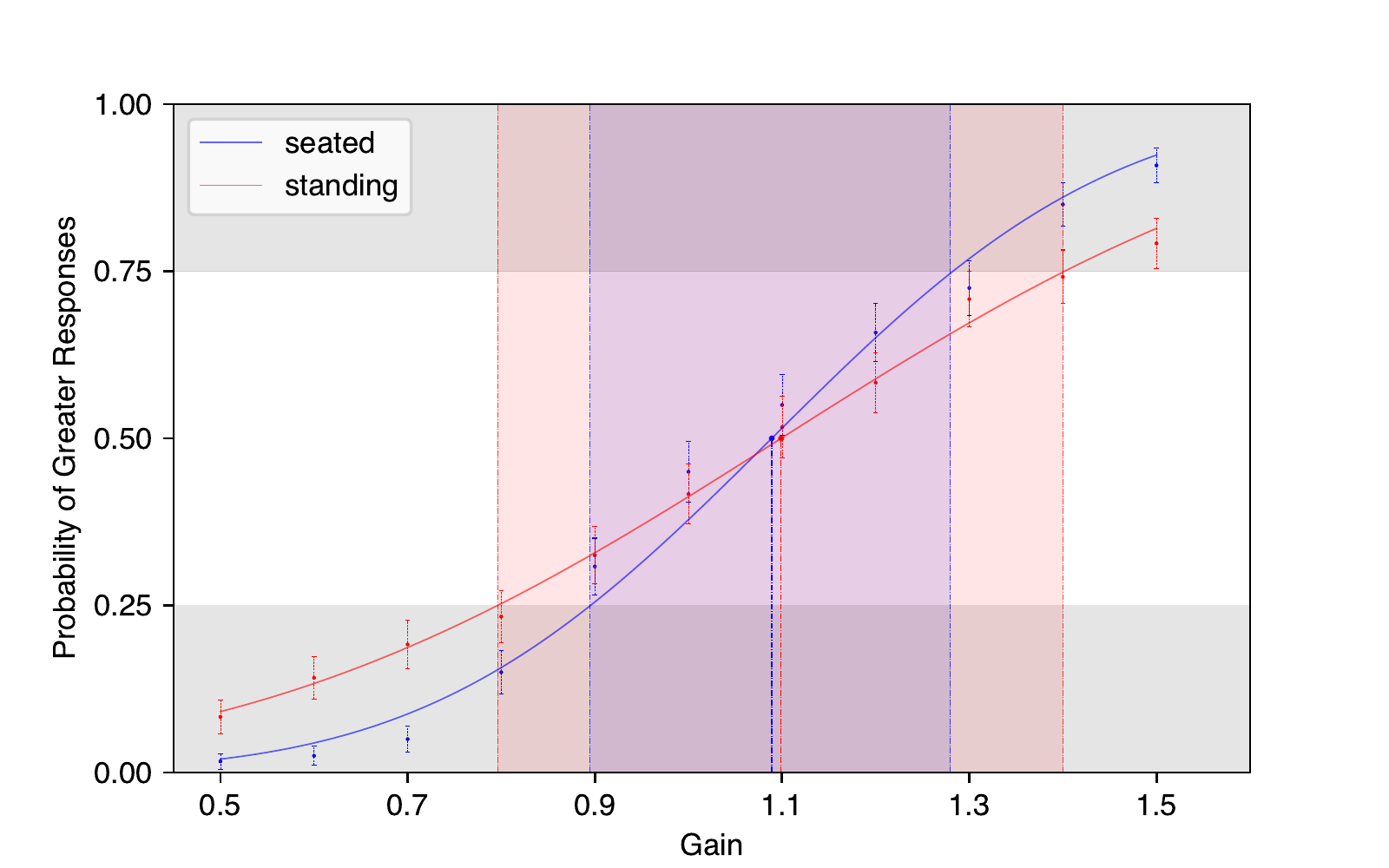}
    \caption{Pooled probability of answering ``greater'' to the 2AFC question for each tested gain and the fitted psychometric curve (error bar shows one standard error). The left and right edges of the colored region show the 25\% and 75\% detection thresholds.}
    \label{fig:curve}
\end{figure}

We calculated the fraction for ``greater'' answers for all scale values with the recorded selections for each trial for both blocks. A psychometric function can thus be fitted with maximum likelihood estimation to model participants' responses to different levels of stimulus \cite{wichmann2001psychometric}. The gain at which users respond ``greater'' with half probability is called the point of subjective equality (PSE), which means they cannot detect stimulus at this level. Detection thresholds were the gains where participants have 25 percent and 75 percent possibilities to give the response of ``greater'', manipulations in between were considered imperceptible to users. We used quickpsy\cite{linares2016quickpsy} (version 0.1.5.1), an R package to perform data analysis that uses the following psychometric function:
\begin{equation}
    \psi(x; \gamma, \lambda, \alpha, \beta) = \gamma + (1 - \gamma - \lambda) * F(x; \alpha, \beta)
\end{equation}
where $\gamma$ and $\lambda$ are the guess rate and the lapse rate that adjusts leftward and rightward asymptotes and can be set to zero if probabilities are close to zero near $x = 0.0$ and $x = 1.0$\cite{benda2020determining}. $F(x; \alpha, \beta)$ denotes a sigmoidal-shape function with asymptotes at 0 and 1. The choice of $F$ has little impact on threshold calculation\cite{linares2016quickpsy}, thus we adopted the \textit{cumulative normal distribution} function as in \cite{williams2019estimation, esmaeili2020detection}:
\begin{equation}
    F(x; \alpha, \beta) = \frac{\beta}{\sqrt{2\pi}}\int_{-\infty}^{x}\text{exp}(-\frac{\beta^{2}(x-\alpha)^{2}}{2})
\end{equation}
where $\alpha$ represents PSE and $\beta$ its standard deviation respectively. Three participants with a probability of fit less than 0.05 were considered to have a bad fit and were excluded from data analysis. Gains at the 25\%, PSE and 75\% were calculated. Fig. \ref{fig:curve} shows the fitted curves and thresholds. We observed a goodness of fit of the psychometric function of $p=0.85$ for seated, and $p=0.99$ for standing.

From the psychometric function, we found a bias for PSE at 1.09 and 1.10 for the seated and the standing block respectively, meaning that most users tend to overestimate head rotations (See Table \ref{tab:gender} for individual results). The 25\% and 75\% threshold is 0.89 and 1.28 for the seated block, a smaller range than for the standing block (0.80 and 1.40). We further compared the thresholds with a 2 genders $\times$ 2 blocks ANOVA. Results demonstrated a significant effect of block, both in 25\% threshold ($F(1,18)=12.015, p=0.003$) and 75\% threshold ($F(1,18)=6.666, p=0.019$), \revision{and no interaction effect was found. This indicates that people are more sensitive to lower and upper bounds of detection thresholds in the seated condition, and gender influence was not found.}


\begin{table}[htbp]
\caption{The 25\%, PSE and 75\% of the psychometric curve for each individual.}
\label{tab:gender}
\centering
\begin{tabular}{l|c|lll|lll}
\hline
& \multicolumn{1}{c|}{\multirow{2}{*}{ID}} & \multicolumn{3}{c|}{Seated} & \multicolumn{3}{c}{Standing} \\
 & \multicolumn{1}{c|}{} & 25\%    & PSE    & 75\%   & 25\%    & PSE     & 75\%    \\ \hline
\multirow{10}{*}{\rotatebox[origin=c]{90}{Male}} & 1  & 0.9723 & 1.0783 & 1.1843 & 0.9069 & 1.0727 & 1.2386 \\
& 2  & 0.8733 & 1.2138 & 1.5544 & 0.5386 & 1.0002 & 1.4617 \\
& 3  & 1.0161 & 1.1958 & 1.3754 & 0.9159 & 1.2707 & 1.6255 \\
& 4  & 0.7875 & 1.2252 & 1.6629 & 0.7982 & 1.0509 & 1.3037 \\
& 5  & 0.9160 & 1.0515 & 1.1871 & 0.5639 & 1.0286 & 1.4933 \\
& 6  & 0.9109 & 1.1169 & 1.3229 & 1.0540 & 1.3777 & 1.7014 \\
& 7  & 0.9452 & 1.0483 & 1.1515 & 0.7068 & 1.0270 & 1.3473 \\
& 8  & 0.8834 & 1.2024 & 1.5214 & 0.7913 & 1.0900 & 1.3887 \\
& 9  & 0.8704 & 0.9717 & 1.0731 & 0.5086 & 0.9711 & 1.4336 \\
& 10 & 0.9509 & 1.0481 & 1.1453 & 1.0098 & 1.1744 & 1.3391 \\
\hline
\multirow{10}{*}{\rotatebox[origin=c]{90}{Female}}  & 11 & 0.8660 & 1.0196 & 1.1732 & 0.7094 & 1.1820 & 1.6546 \\
& 12 & 0.9622 & 1.1390 & 1.3158 & 0.6841 & 1.1428 & 1.6014 \\
& 13 & 0.8023 & 0.9591 & 1.1160 & 0.7774 & 1.0611 & 1.3447 \\
& 14 & 0.6928 & 1.0227 & 1.3526 & 0.7498 & 1.3140 & 1.8781 \\
& 15 & 0.7792 & 1.0479 & 1.3166 & 0.7654 & 1.0831 & 1.4008 \\
& 16 & 1.0244 & 1.2222 & 1.4199 & 0.9824 & 1.1240 & 1.2656 \\
& 17 & 0.8623 & 1.0362 & 1.2101 & 0.8637 & 0.9613 & 1.0589 \\
& 18 & 0.9819 & 1.1169 & 1.2519 & 0.8450 & 1.1133 & 1.3816 \\
& 19 & 0.9508 & 1.0724 & 1.1939 & 0.7604 & 1.0152 & 1.2700 \\
& 20 & 0.9931 & 1.1384 & 1.2838 & 0.8110 & 1.0818 & 1.3526 \\
\hline
\multicolumn{2}{c|}{Male}  & 0.9040  & 1.1005 & 1.2971 & 0.7892  & 1.1081  & 1.3782  \\ \hline
\multicolumn{2}{c|}{Female} & 0.8856  & 1.0766 & 1.2676 & 0.8027  & 1.0905  & 1.4270  \\ \hline
\multicolumn{2}{c|}{All} & 0.8947  & 1.0885 & 1.2823 & 0.7961  & 1.0991  & 1.4020  \\ \hline
\end{tabular}
\vspace{1pt}
\end{table}

\subsubsection{Response Time}
\begin{figure}[htbp]
    \centering
    \includegraphics[width=0.9\linewidth]{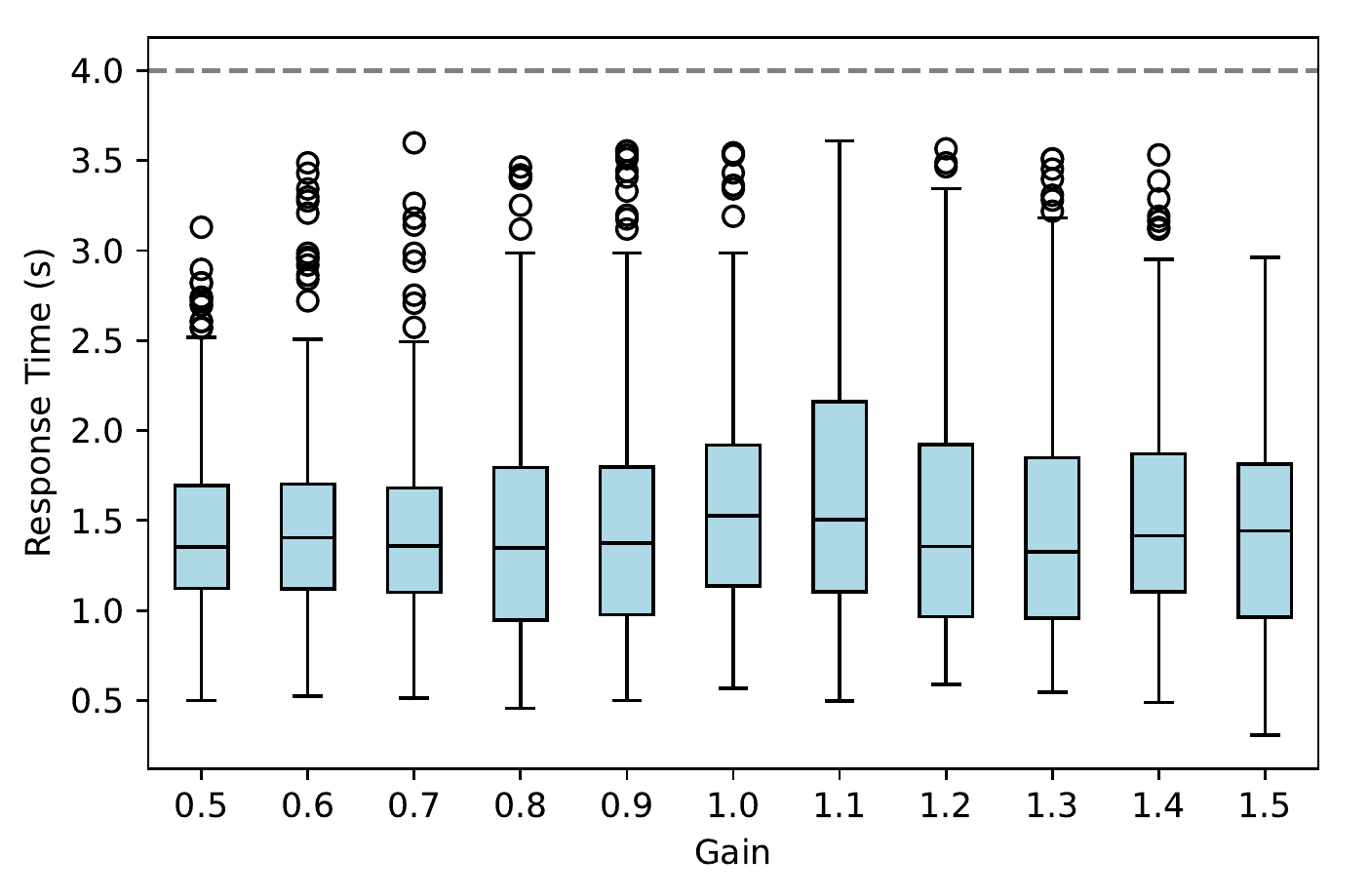}
    \caption{Boxplot of response time for each rotation gain.}
    \label{fig:response_time}
\end{figure}

With the logged response time for each trial, we removed outliers following the 1.5 interquartile rule and boxplotted the data (See Fig. \ref{fig:response_time}). On average, participants reacted timely and gave their answers within four seconds. Times were the longest with gains \revision{at 1.0 and 1.1}, which indicates that they caused confusion to users. \revision{On the other hand, it took the shortest time for participants to decide at gains 1.5.} We performed a comparison of response time with an ANOVA considering 11 gains $\times$ 2 genders $\times$ 2 blocks. There were significant main effects of gain ($F(10, 180) = 3.611, p < 0.001$) and block ($F(1, 18)=7.458, p=0.014$), a trending effect of gender ($F(1, 18) = 3.950, p = 0.062$). There is also an interaction effect between gender and gain ($F(10, 180) = 2.345, p = 0.013$).

\subsubsection{Simulator Sickness}
In order to measure simulator sickness, we calculated the SSQ total severity (TS) score. We measured a mean SSQ-score of 6.36 ($SD = 6.68$) and 27.87 ($SD = 29.26$) before and after the seated block, 4.30 ($SD = 4.52$) and 30.48 ($SD = 32.00$) before and after the standing block. A Kolmogorov-Smirnov test showed that the data was not normally distributed. We analyzed the results with a Wilcoxon signed-rank test at the 5\% significance level and found that the SSQ score was significantly higher after experiencing the seated 
block ($p = 0.0004$) and the standing block ($p=0.0003$). ANOVA found no significant effect of block ($F(1,18)=0.505, p=0.486$) or gender ($F(1,18)=1.199, p=0.288$) on the SSQ-score change as well as any interaction effect ($F(1,18)=0.554, p=0.466$). 

\subsection{Discussion}
Our results indicate that participants are unable to discriminate a real rotation deviating from a fixed virtual rotation of 90$^{\circ}$ between $-11.8\%$ and $+22.0\%$ while seated. In the standing condition, real rotations can deviate between $-25.6\%$ and $+28.7\%$, which indicates that users are more sensitive to rotation modification in stationary conditions with restricted body movements. The results are quite interesting considering the movement cues that participants received. In the seated block, they mainly relied on visual information for the virtual environment, while for the real world they sensed their head rotation through the vestibular system and lacked proprioceptive cues from the upper body compared to standing condition. If only vestibular and visual cues were presented, the subordinate proprioceptive cue is removed while the dominant cues remained (vestibular stimulation dominates over other sensory cues of motion\cite{borah1979sensory, williams2019estimation}). It would be easier for users to tell the difference between real and virtual motion with less conflict. On the other hand, the relative movement between the head and upper body in the seated block also makes real-world head movements more palpable to participants.

Our thresholds for the standing condition are tighter than those in Williams and Peck's work \cite{williams2019estimation} that also tested on a $110^{\circ}$ FOV. A possible reason is that for a fair comparison, we set the same range of head rotation angles in both blocks considering the maximum angles users could rotate while seated. With limited head rotation, participants received less proprioceptive cues than in previous work. This also suggests that even when users stand, the manipulation should be more conservative with applications that involve mostly forward interactions.

The results in the seated block update the previous findings under the condition of controlled upper body movement by Jaekl et al. \cite{jaekl2005perceiving} using a state-of-the-art device. Our experiment further reaffirms that there is a trade-off between less physical movements and sense of naturalism. With only head motion, the gains are more susceptible to be noticed compared to standing. Therefore, HMD designers and application developers need to be more cautious to apply rotation gains in seated VR situations. 

We did not find a significant difference between two genders as in \cite{williams2019estimation}. However, the age range of male participants in their experiment has much larger deviance. We cannot rule out the possibility of age influencing detection thresholds. The difference might also be explained by individual differences. We found PSE varies largely among participants ($0.97$ to $1.22$ and $0.96$ to $1.37$ for seated and standing) and the discrepancy \cite{williams2019estimation} is even larger.

Head rotation within perceptual limitations allows users to interact with virtual content naturally. However, according to our results, the upper detection threshold for seated VR is even smaller than previously estimated for redirected walking. When being stationary seated, users can only explore $1.28 \times 90 \times 2 = 230$ degrees even if they rotate their head from $-90$ degree to $90$ degree. This would be already at their physical limit and much larger than their \textit{maximum comfortable rotation angle}, but still far from sufficient for 360-degree of free viewing.
\section{Experiment 2: Applicability}
In the previous section, we analyzed the thresholds for detecting an amplified rotation. However, when looking into aspects of applicability it is more important to explore how a complete 360 scene with limited head motion can be experienced. In this case, it is more reasonable to regard an amplified head rotation as an interaction technique rather than confine it within the undetectable range. Earlier work has demonstrated that rotations gain can be used beyond perceptual limitations \cite{sargunam2017guided, ragan2016amplified, langbehn2019turn}. However, these works performed either evaluation within a specific task or comparison with other non-linear mappings. Freitag et al.\cite{freitag2016examining} classified rotation gains into \textit{negligible}, \textit{tolerable} and \textit{unfit for continuous use} in CAVE system, which also motivates us to explore user acceptance of rotation gains in sitting scenarios. To the best of our knowledge, there is no work on users' reactions to higher levels of gains and how much gain, even if it can be noticed, could be accepted by users when they are seated. 

To address this gap, we designed and conducted our second experiment. We used a set of items proposed by Rietzler et al. \cite{rietzler2018rethinking} to measure the applicability of a gain. The first item is naturalness. It describes how much participants feel the rotation has deviated from the most natural mapping. The second item combines nausea and sickness and reflects the assumption that high gain will possibly introduce more sense of disorientation and motion sickness. 
We decided against using comfort as an item as Rietzler et al.'s results \cite{rietzler2018rethinking} in the boxplots showed that the distribution of \textit{pleasant} and \textit{applicable} was highly consistent. Instead we included physical movement as a third item since in seated VR, the amount of physical movement required for different gains have a great discrepancy. Note that physical movement cannot be properly estimated from head rotation data because users need to move their upper body at least in a one-to-one mapping. The fourth item we used for measuring applicability is applicability of a gain itself and directly reflects users' willingness to use a gain for seated VR applications \cite{rietzler2018rethinking}.

\subsection{Design}

In our experiment, we analyzed the preferences of users towards rotation gains starting from 1.0 with a step of 0.5. We set the upper bound to 3.5 by which users would only need 51-degree head rotation for 180-degree viewing, an angle smaller than the maximum comfortable rotation angle of most users. We divided the experiment into two blocks to investigate possible adaptation effects. We randomized gains and tested twice in both blocks and participants had to accomplish 2 practice + 2 blocks $\times$ 6 gains $\times$ 2 repeats = 26 trials. 

\begin{figure}[htbp]
    \centering
    \includegraphics[width=0.9\linewidth]{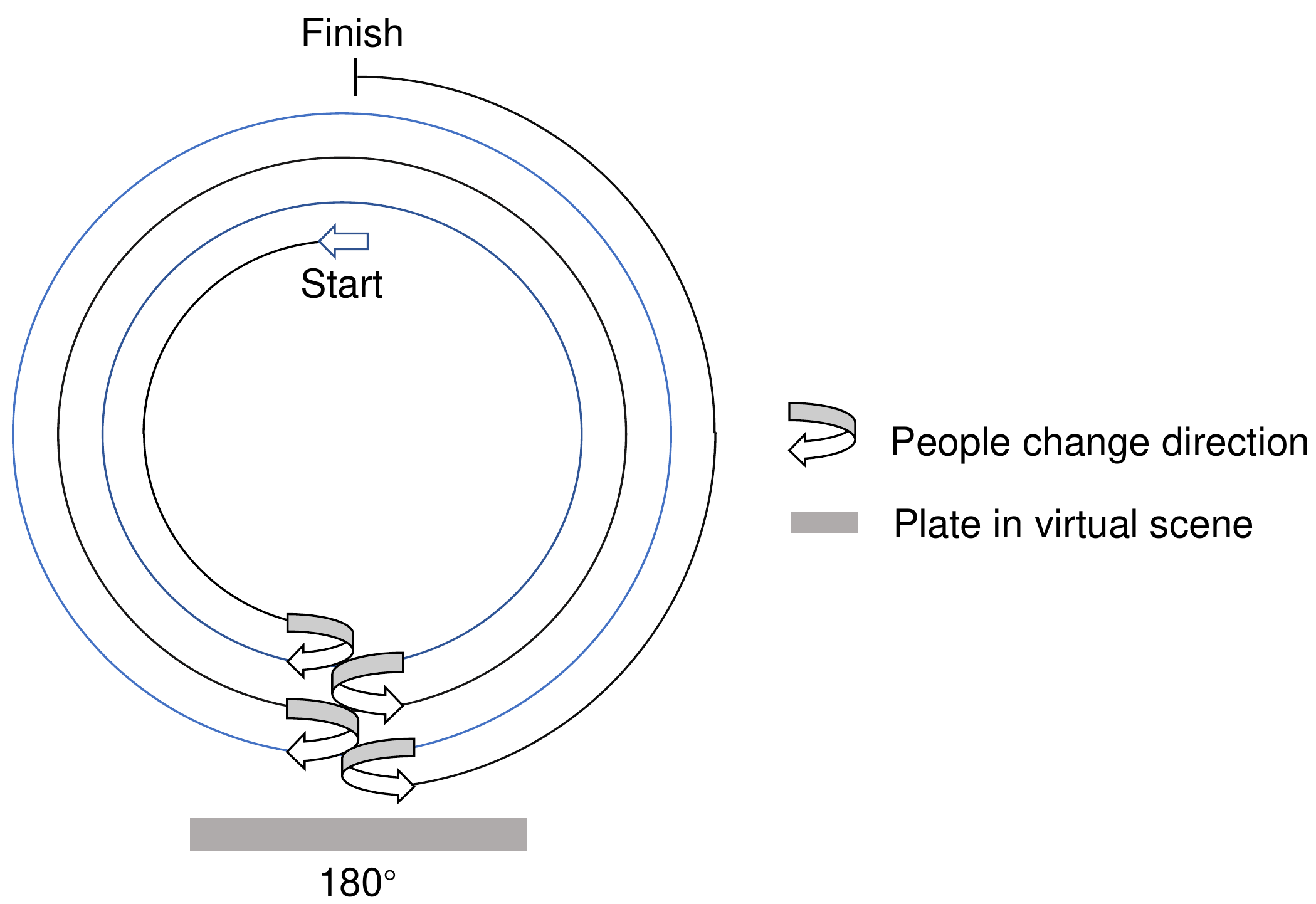}
    \caption{The rotation process of the second experiment in the virtual scene when the initial direction was left. Participants started from the innermost half-circle, rotated in the direction of the start arrow, they changed their directions when hitting the plate and rotated according to the second smallest circle. They followed the same strategy for the remaining circles. The pattern looks similar if the initial direction was right.}
    \label{fig:exp2-process}
\end{figure}

One of the goals for our experiments was to make it independent of a specific task avoiding that participants might be influenced by task-specific actions and that the results may fail to be generalized. Thus we used a 360-degree free viewing task that did not include any time pressure. We placed a white plate in the virtual world located at 180-degree. Once hit by the user this would turn to green. Each trial contained a 360 $\times$ 4 = 1440 virtual degree rotation in a continuous process (See Fig. \ref{fig:exp2-process}) and would provide enough exposure for a certain gain. The virtual rotation was the same for each gain for a fair comparison. Participants initiated their head rotation with the direction marked by an arrow until the plate appeared in the center of their eyes. They then rotated in the opposite direction to hit the plate again three times. Afterwards, they rotated back to the neutral position and questions would be displayed. We used a seven-point Likert scale questionnaire to collect user feedback to the four statements listed below (1 means totally disagree and 7 means totally agree).
\begin{itemize}
    \item Rotating the head like this in a virtual world is natural.
    \item Rotating the head like this in a virtual world requires a lot of physical movements.
    \item I have a strong feeling of nausea or sickness in the trial.
    \item Rotating my head like this in a VE is applicable. 
\end{itemize}

We did not explicitly control the head rotation speed of participants but we suggested that participants rotate with a normal and comfortable speed to see the scene clearly. Furthermore, participants were told to not move their upper body unless necessary (this happened at gains of 1.0 and 1.5 for most participants). A possible consideration is that participants might infer the rotation gain from the position of the plate, but it is less a problem. First, we did not point out the exact virtual position of the plate in our instructions and the virtual scene would be rotated in each trial, making it hard to deduce how large the gain was. Second, the question should be answered based on the subjective experience during a trial regardless of the value of the gains.

\subsection{Participants}
We recruited 13 participants from the campus  (mainly students and employees of Tsinghua University, none of them participated in the first experiment). The participants were aged between 19-27 (6 male and 7 female, mean age 21.3, $SD=2.12$) and all successfully accomplished the study. All participants had a normal or corrected-to-normal vision (9 wear glasses or contact lenses) and were physically and mentally healthy. Among them, 3 were novice VR users who never experienced VR before, 6 had at least some experience with VR and 4 rated themselves as frequent VR users. 6 of them reported extensive video game experience. 

\subsection{Procedure}
The apparatus was similar to the first experiment. We obtained informed consent from participants before the beginning of the experiment. Each of them received \$12 for compensation. Then participants would be briefed about the goal and the process of the experiment. The experimenter later helped participants to adjust (the IPD and band length) and to wear the HMD after they finished a demographic survey and the SSQ questionnaire. Then two practice trials would start after which participants could ask questions about the procedure. After each trial, the four questions mentioned above would be displayed and participants were asked to answer them with a controller. We logged answers for further analysis. All participants finished the experiment with the Oculus Touch right-hand controller. Throughout the experiment, we strongly advised participants to take a 10-20 second break after any trial to mitigate the accumulation of sickness. Participants could also choose to take longer breaks up to two minutes. After the practice run, the first block would start. A 5-minute rest was enforced between the first and second block during which participants filled another SSQ. Afterward, they proceeded to the second block. The third SSQ would be distributed once all trials were completed. The entire procedure took about 60 minutes.

\subsection{Results}
We collected the results and further analyzed them using the appropriate statistical tests.
\subsubsection{Applicability items}
\begin{figure}[htbp]
    \centering
    \includegraphics[width=0.95\linewidth]{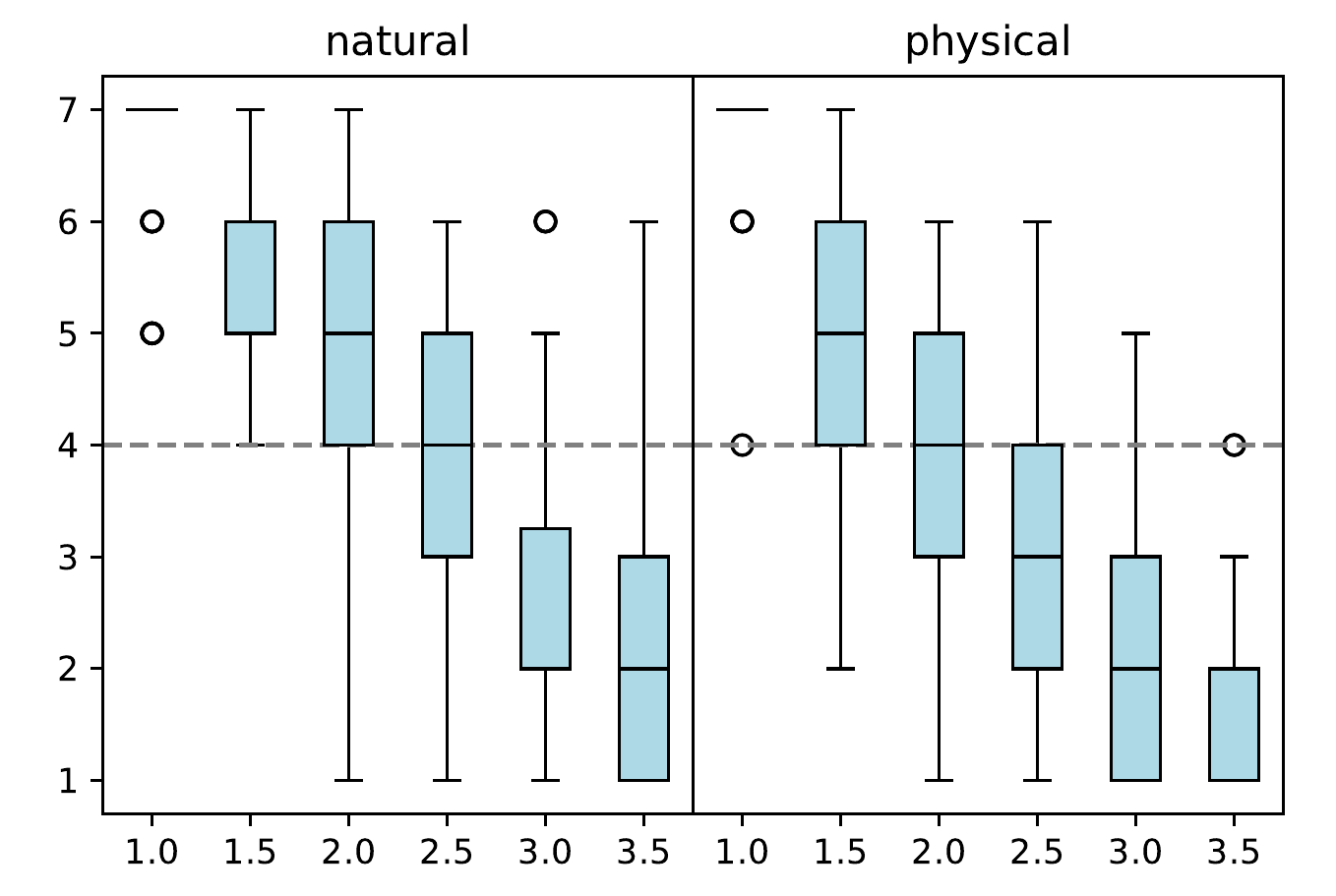}
    \includegraphics[width=0.95\linewidth]{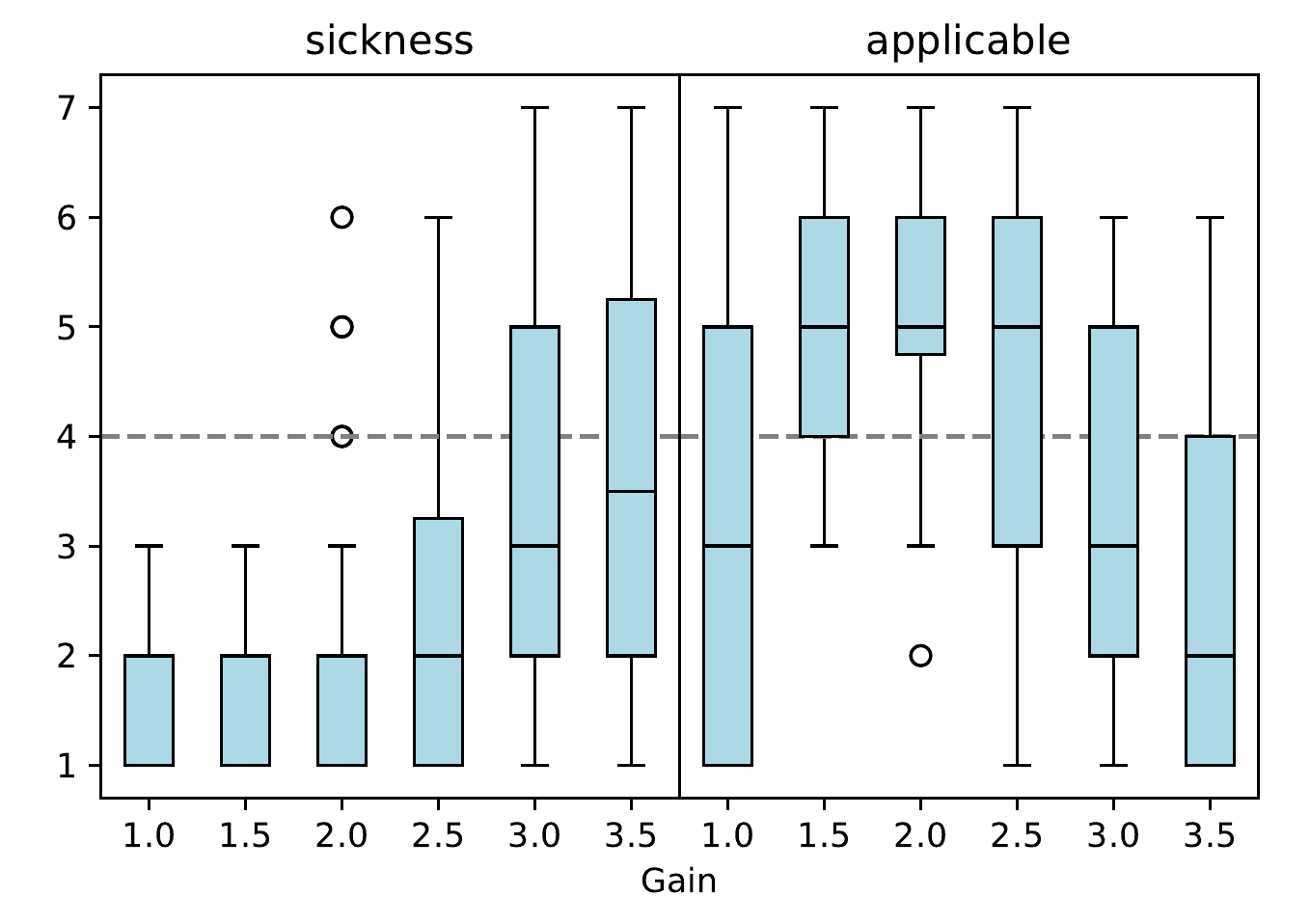}
    \caption{Boxplots for each applicability item, the x-axis represents the different gains, the y-axis represents the 7-Likert scale.}
    \label{fig:bp}
\end{figure}

A Friedman test indicated a statistically significant difference in each of the items depending on the different gains (Fig.\ref{fig:bp}, $\chi^{2}(5) = 55.752, p < 0.0001$ for naturalness, $\chi^{2}(5) = 61.991, p < 0.0001$ for physical, $\chi^{2}(5) = 34.068, p < 0.0001$ for sickness, $\chi^{2}(5) = 30.798, p < 0.0001$ for applicable). We further used a Wilcoxon signed-rank test with Bonferroni correction for multiple pairwise comparisons and reported significance at 5\% level.

\textit{Naturalness}. There were significant differences between 1.0 and other gains ($p < 0.01$). Users could easily notice the manipulation was beyond the threshold if the gain was not 1.0. We did not measure a significant difference between 1.5 and 2.0 ($p = 0.12$), but naturalness continued to diminish significantly when compared both of them with gains larger than 2.0 ($p < 0.05$). 

\textit{Physical}. Similar to naturalness, the physical movement answers between a gain of 1.0 and others were significantly different ($p < 0.001$ for all tests). For seated VR, gains equal or greater than 1.5 could effectively reduce physical motion.
The differences between 1.5 and 2.0, 2.0 and 2.5 were not significant ($p = 0.845, 1.00$), however, 1.5 differs significantly from 2.5 and larger ($p < 0.05$ for all tests), 2.0 differs significantly from 3 and larger ($p = 0.004, 0.003$). For gains greater than 2.5, users did not feel that their movements in the virtual world require a lot of physical movement  ($p > 0.05$ for tests between any of 2.5, 3.0 and 3.5). Users only needed to rotate their heads up to 72 degrees for a gain of 2.5, close to the maximum comfortable rotation angle. Hence, gains exceeding 2.5 seem to not influence the feeling of decreased body motion to a substantial extent.

\textit{Sickness}. The sickness scores were only significantly different when comparing gains of 1.0 to 3.0 and 3.5 ($p = 0.018, p=0.020$), as well as 1.5 to 3.0 and 3.5 ($p = 0.026, p=0.020$).

\textit{Applicability}. Applicability differed significantly between gains of 1.5 and 3.0 ($p = 0.027$), 1.5 and 3.5 ($p = 0.008$), 2 and 3 ($p = 0.016$), 2 and 3.5 ($p = 0.004$), 2.5 and 3.5 ($p = 0.024$).

\begin{figure}[htbp]
    \centering
    \includegraphics[width=0.95\linewidth]{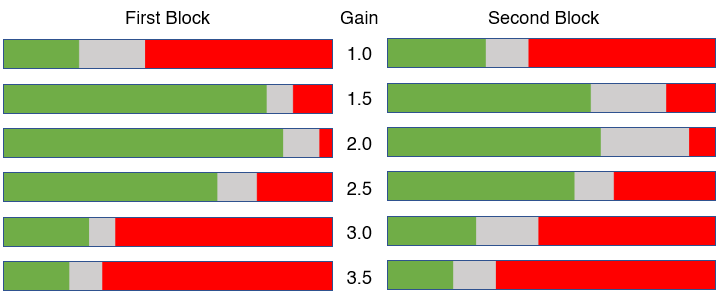}
    \caption{Percentage of whether a gain was regarded as acceptable (green), neutral (gray) or unacceptable (red) in two blocks.}
    \label{fig:percent}
\end{figure}

We also plotted the proportion of acceptance for each gain in both blocks (Fig. \ref{fig:percent}). Applicability with scores greater than 4, equal to 4 and less than 4 were considered acceptable, neutral and unacceptable respectively. The three most applicable gains were 2.0, 1.5 and 2.5. However, their acceptances in the second block decreased, possibly due to the increase of sickness levels.

\subsubsection{View Time}
\begin{figure}[htbp]
    \centering
    \includegraphics[width=0.9\linewidth]{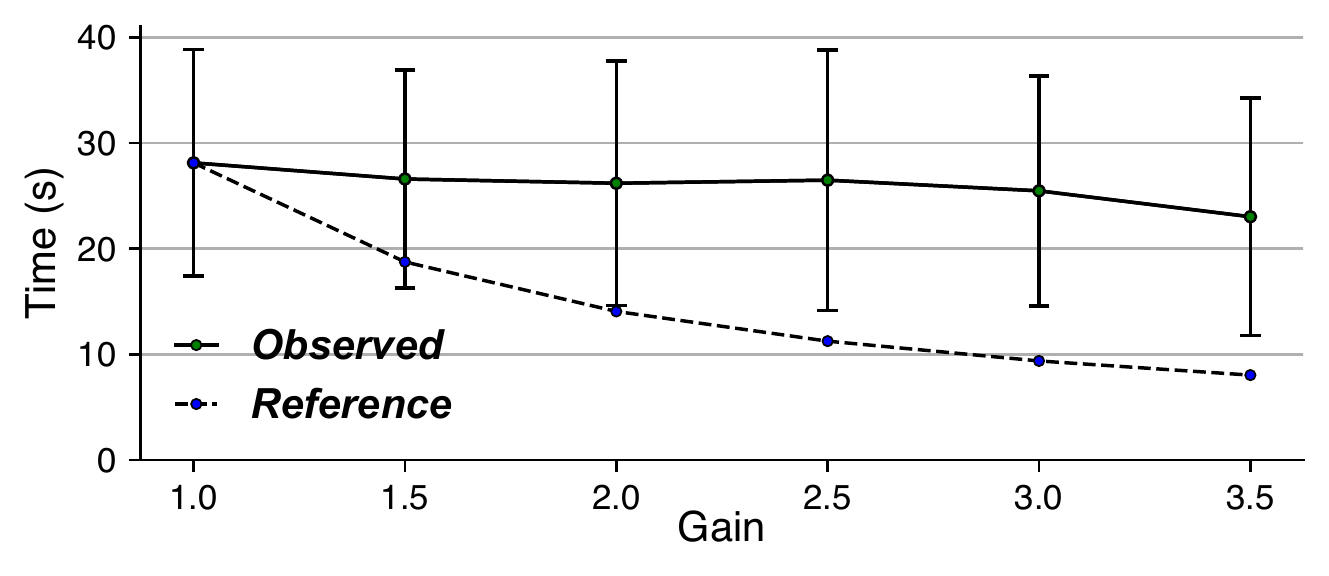}
    \caption{View Time for each gain when observed in our experiment (\textit{Observed}) or when participants keep the same head rotation speed as the gain is 1.0 (\textit{Reference}). For the \textit{Observed} line, the error bar shows one standard deviation. See how participants adjust their head rotation speed according to the gain and cost more time than reference.}
    \label{fig:view-time}
\end{figure}

The view time reveals how long participants experienced a gain, referring to the duration between the time they started rotating their head and the questions appeared. Participants spent on average 20 to 30 seconds for every gain, only slightly longer for smaller gains (See Fig. \ref{fig:view-time}). It was interesting to note that users tended to alter their head rotation speeds according to different gains so that they traversed the virtual scene with a similar speed.

We analyzed the view time with an ANOVA considering 6 gains $\times$ 2 blocks $\times$ 2 genders. A significant difference was found of block ($F(1, 24) = 4.649, p  = 0.0319$) and gender ($F(1, 24) = 16.155, p < .0001$). Users rotated faster in the second block and female participants spent more time than male participants. No other interaction effects were found.

\subsubsection{Simulator Sickness}
\begin{table}[htbp]
\caption{SSQ score (Mean $\pm$ SD) in each stage.}
\label{tab:ssq-exp2}
\centering
\begin{tabular}{ccc}
\hline
Before & Middle & After \\ \hline
$2.181 \pm 5.398$ & $17.765\pm19.222$ & $25.557\pm20.412$ \\ \hline
\end{tabular}
\vspace{2pt}
\end{table}

We measured the SSQ scores before the experiment, during the middle break and after the all trials (Table \ref{tab:ssq-exp2}). A Wilcoxon signed-rank test with a 5\% significance level found that SSQ score in the middle was significantly higher than that at the start ($p = 0.002$). The final SSQ score was also significantly higher than the middle ($p = 0.036$).

\subsection{Discussion}
Rotation gains of 1.0 caused the least sickness and were still considered to be the most natural. In redirected walking techniques participants often favored no additional manipulation (e.g. \cite{rietzler2018rethinking}), but this is not the case for seated VR head amplification in which physical movement is also an essential concern. To fully explore the virtual scene, the most natural mapping required users to move their bodies extensively. Actually, there is a balance between physical turning and naturalness along with sickness. We found that participants have diverse opinions about a gain of 1.0. Three participants regarded it as applicable, among them two gave the highest applicability score, while other participants rated it as not applicable for seated VR.

Most users preferred manipulation that made them move less. From the results, we can see that a gain of 2.0 was rated the best in terms of applicability and about 60\% participants would accept rotation gains at 2.5, far exceeding the detection thresholds. A gain of 2.5 is close to the verge of a usable gain since it does both, it reduces movements and seems to not increase nausea as we could not measure a significant difference compared to a gain of 1.0. Furthermore, 27\% and 19\% of the subjects were even comfortable with gains at 3.0 and 3.5. However, we cannot neglect the fact that in the second block, users lowered their ratings for large gains. This is most likely due to their cybersickness increased with time (see the SSQ score in Table \ref{tab:simulation}) especially with the rotation gains changed from trial to trial. In the second block, they leaned toward more natural gains.

Therefore, we have affirmed that head amplification can be treated as an interaction technique and noticeable gains are usable for seated VR. Though gains around 2.5 were applicable for many participants, they may not be suitable for long time usage and should be examined carefully for different users. We particularly highlight the importance of an individual setup. One may either favor natural rotation or tend to be ``lazy'' and not willing to move, or even a single user can be versatile and have contradicting preferences over time. Moreover, now that users are already aware of the manipulation, the design can enable them to alter the gain instantly to their current need, which might prolong the available time for large gains and avoid possible adverse side effects. If the region of interest is already known, it is ideal to combine amplification with redirection as in \cite{sargunam2017guided} to reduce the time to constantly rotate around and be exposed to higher gains. Our results provide reference to the range of gain possible for the amplification phase. It is also possible to combine large rotation gains with other techniques such as viewpoint snapping that reduce continuous viewpoint motion by skipping frames to reduce cybersickness \cite{farmani2018viewpoint}.


Head rotation speed is also an interesting aspect worth discussing. Although participants were advised to maintain a normal speed during the experiment, they could actually adjust it because the experiment design did not strictly enforce the speed. If they rotated with the same speed regardless of gains, the amount of visual information received per time frame for rotation gain at 3.0 would be three times than that of 1.0, making users more prone to sickness and making it hard to see the virtual scene clearly. Our results showed that they would instead rotate slower with larger gains to assure the influx of optical flow did not fluctuate too much. Whereas not the main goal of our study, it is possible to further investigate users' sensitivity and acceptance to rotation gains with controlled head rotation speeds. The relationship has already been established in redirected walking \cite{neth2012velocity}.

\section{Simulation}
We did not find a significant difference in the amount of physical movement between a gain of 2.0 and 2.5. Both gains enabled users to traverse the 360-degree scene with only head movements and are an option for practical use. Hence, we performed a computer simulation to further explore how the movement may differ in more complex situations. We selected a visual searching task that involved enough head-turning: users needed to rotate their head to hit $n$ balls (here hit means the head yaw angle is equivalent to the ball rotation). In practice, the ball might be a target in a game, or an interesting character in a movie. At any time, only one ball would appear at the scene and when a ball is hit another one would be randomly generated elsewhere. We experimented with several levels of restrictions on the distance between two continuous balls (no restrictions or at least some degrees apart) considering that balls might be compact or scattered. The parameter $n$ was empirically set to 10. For simplicity, we assumed that users would always rotate in the right direction of the next ball. We compared a constant gain of 2.0, 2.5 and other two non-linear methods that also enabled 360-degree within 180-degree head rotation ($2-cos(\theta)$ in \cite{sargunam2017guided}, quadratic in \cite{langbehn2019turn}). Each restriction pattern was repeated 1,000,000 times. We calculated how many angles were needed for the head rotation and beyond maximum comfortable rotation angle (set to 60 degree). The results are shown in Table \ref{tab:simulation}.

\begin{table}[htbp]
\caption{Simulated physical movements in degrees for each method (values in brackets are movements beyond the maximum comfortable rotation angle). X+ in the leftmost column indicates that two continuous balls are at least X degrees apart.}
\centering
\begin{tabularx}{\linewidth}{ccccc}
\hline
& 2.0 & 2.5 & $2-cos(\theta)$ & Quadratic
\\ \hline
0+ & 451.36 (50.7) & 361.09 (10.3) & 555.15 (80.6) & 451.35 (39.6) \\ \hline
30+ & 454.26 (52.2) & 363.40 (10.8) & 557.43 (81.9) & 454.07 (41.0) \\ \hline
60+ & 464.02 (56.2) & 371.22 (11.8) & 565.55 (86.4) & 462.84 (44.3) \\ \hline
90+ & 481.84 (61.5) & 385.47 (12.9) & 581.11 (93.7) & 478.59 (48.6) \\ \hline
\end{tabularx}

\label{tab:simulation}
\end{table}

Generally, a gain of 2.5 required the least physical motion. A gain of 2.0 and quadratic were almost the same, and they were much less than $2-cos(\theta)$. As for the movement beyond the maximum comfortable rotation angle, the ordering was gain at 2.5, quadratic, gain at 2.0, $2-cos(\theta)$ from less to more. We found whether the function of the tracked head angle and virtual angle was concave or convex contributed greatly to the result. $2-cos(\theta)$ was completely concave between 0$^{\circ}$ and 90$^{\circ}$, so it was expected to have more physical rotations given the same virtual ball distribution. Quadratic was concave from 0$^{\circ}$ to 45$^{\circ}$ and convex from 45$^{\circ}$ and 90$^{\circ}$, hence the overall movement was similar to that of a gain at 2.0, but movement beyond the maximum comfortable rotation angle was much less. We admit that the difference was accumulated from massive trials, it might not be notable in actual use. However, it still told us that concavity was a crucial consideration when designing a suitable remapping function or strategy.

\section{Limitation and Future Work}
Our first two experiments suffer from the problem of limited sample size and age distribution due to pandemic issues. This might introduce a bias in our results. Another issue is that in this paper, participants only rotated their heads while in reality a small amount of upper body movement is also acceptable even when seated. This needs to be studied more deeply. In the second experiment, though we provided enough time for participants to experience a certain gain, it was far from a real application. Our results may only serve as an upper bound and might not be used constantly. Further research can focus on experiments that enforce longer exposures to large gains and in more practical applications. In this context, controlling the amount of cybersickness to an affordable range might be a major concern. Also, the influence of amplification on spatial orientation was not examined in our work, but we can be safe to conclude that it would worsen as gains became larger. If the virtual scene involves a lot of traveling or is not familiar to users, designers should be more conservative for amplification as users may be confused in their orientation. Future work could also examine the difference of spatial orientation, presence or user performance for these large gains according to related methodologies in \cite{langbehn2019turn, ragan2016amplified, freitag2016examining}. Another factor we did not explore is scene complexity. The scene complexity will have an impact on the amount of optical flow users would experience during rotation. It could be that they might accept large gains with simple scenes. We left it for future work as the complexity was hard to distinguish and should be addressed by delicate experiment design. \revision{Additionally, conducting our experiments under controlled head rotation motion is also a promising direction as it influences the perception of users toward rotation gains~\cite{brument2021studying}. }

Regarding the simulation, we only examined four types of mappings but it is possible to extend it to dynamic rotation gains (e.g. time dependent~\cite{zhang2013human}, velocity dependent~\cite{zhang2021velocity}, or \revision{different rates of change~\cite{congdon2019sensitivity}}), pitch and roll direction and be combined with other seated VR interaction techniques. It can also be used to verify the effectiveness of proposed redirection strategies and guide the design of user studies.
\section{Conclusion}
The primary goal of our research was to investigate human perception of altered head rotation in seated VR. We first conducted a psychophysical experiment using a 2AFC design to estimate detection thresholds of rotation gains in a sitting and a standing condition. We found that users were more sensitive to rotation gains while they are seated compared to when they are standing with whole body rotation. We further performed another experiment to analyze if they could accept gains beyond the threshold and confirmed that suitable gains, even noticeable, could substantially decrease the extensive movement needed by one-to-one mapping while introducing little unnaturalness and sickness.

Our research provides insights about using rotation amplification as an interaction technique for seated VR. Gains larger than thresholds can be used to modify the head movements in VR based on the preference of individuals and the purposes of applications.


%





\ifCLASSOPTIONcaptionsoff
  \newpage
\fi



%



\bibliographystyle{abbrv-doi}

\bibliography{ref}

%

\begin{IEEEbiography}[{\includegraphics[width=1in,height=1.25in,clip,keepaspectratio]{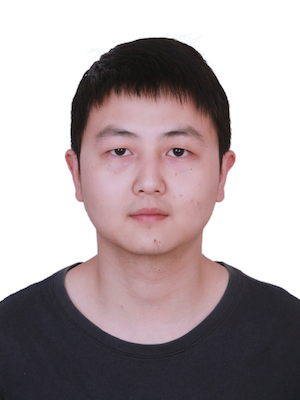}}]{Chen Wang} 
received his bachelor's degree in computer science from Department of Computer Science and Technology, Tsinghua University, where he is pursuing his master's degree. His research interests include computer graphics and computer vision.
\end{IEEEbiography}

\begin{IEEEbiography}[{\includegraphics[width=1in,height=1.25in,clip,keepaspectratio]{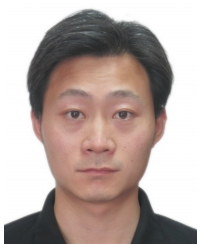}}]{Song-Hai Zhang} received the PhD degree of Computer Science and Technology from Tsinghua University, Beijing, in 2007. He is currently an associate professor in the Department of Computer Science and Technology at Tsinghua University. His research interests include virtual reality and image/video processing. 
\end{IEEEbiography}

\begin{IEEEbiography}[{\includegraphics[width=1in,height=1.25in,clip,keepaspectratio]{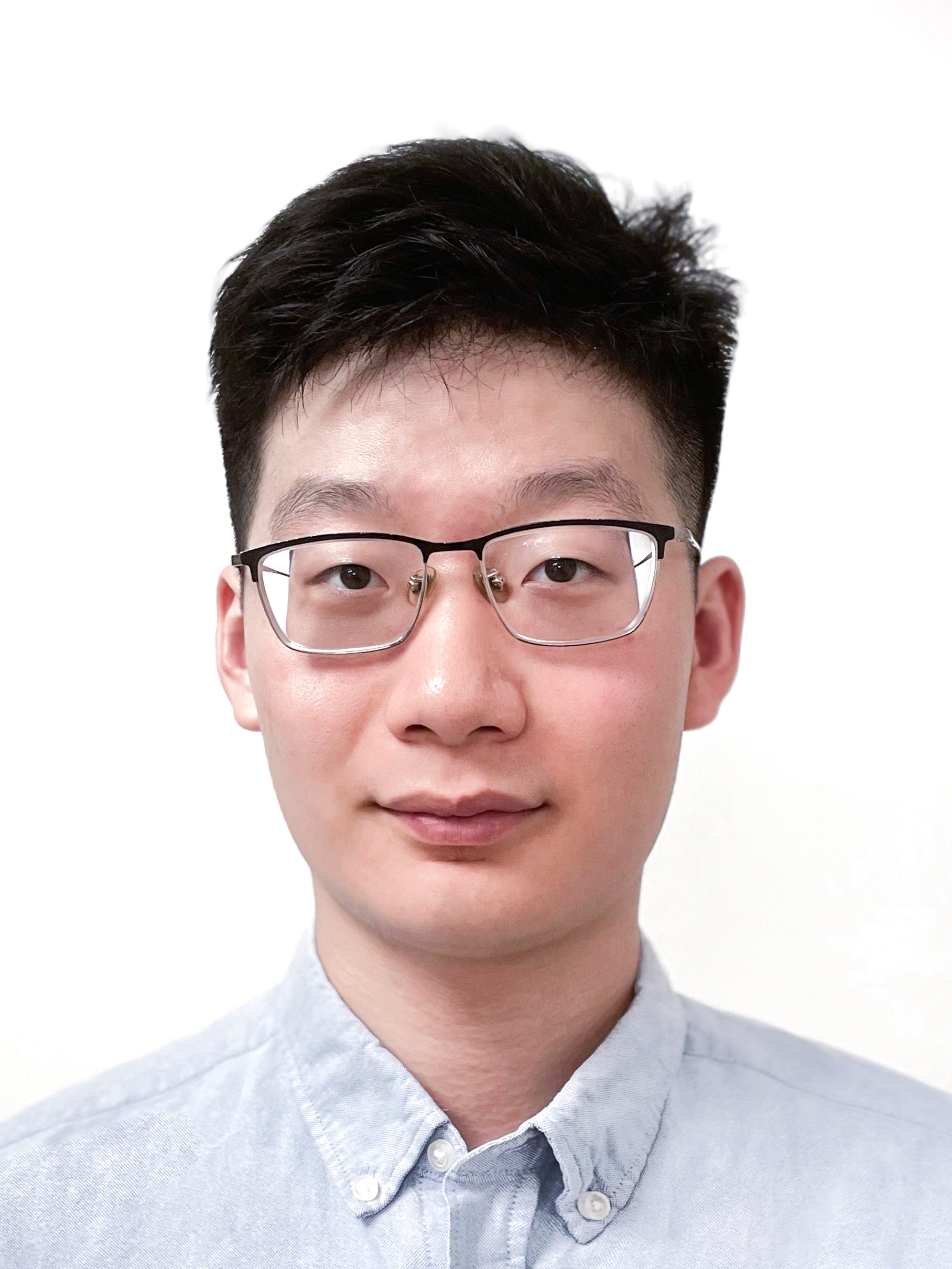}}]{Yizhuo Zhang} is an undergraduate who majors in computer science and technology at Tsinghua University. His research interests include machine learning, human-computer interaction and mixed reality. 

\end{IEEEbiography}

\begin{IEEEbiography}[{\includegraphics[width=1in,height=1.25in,clip,keepaspectratio]{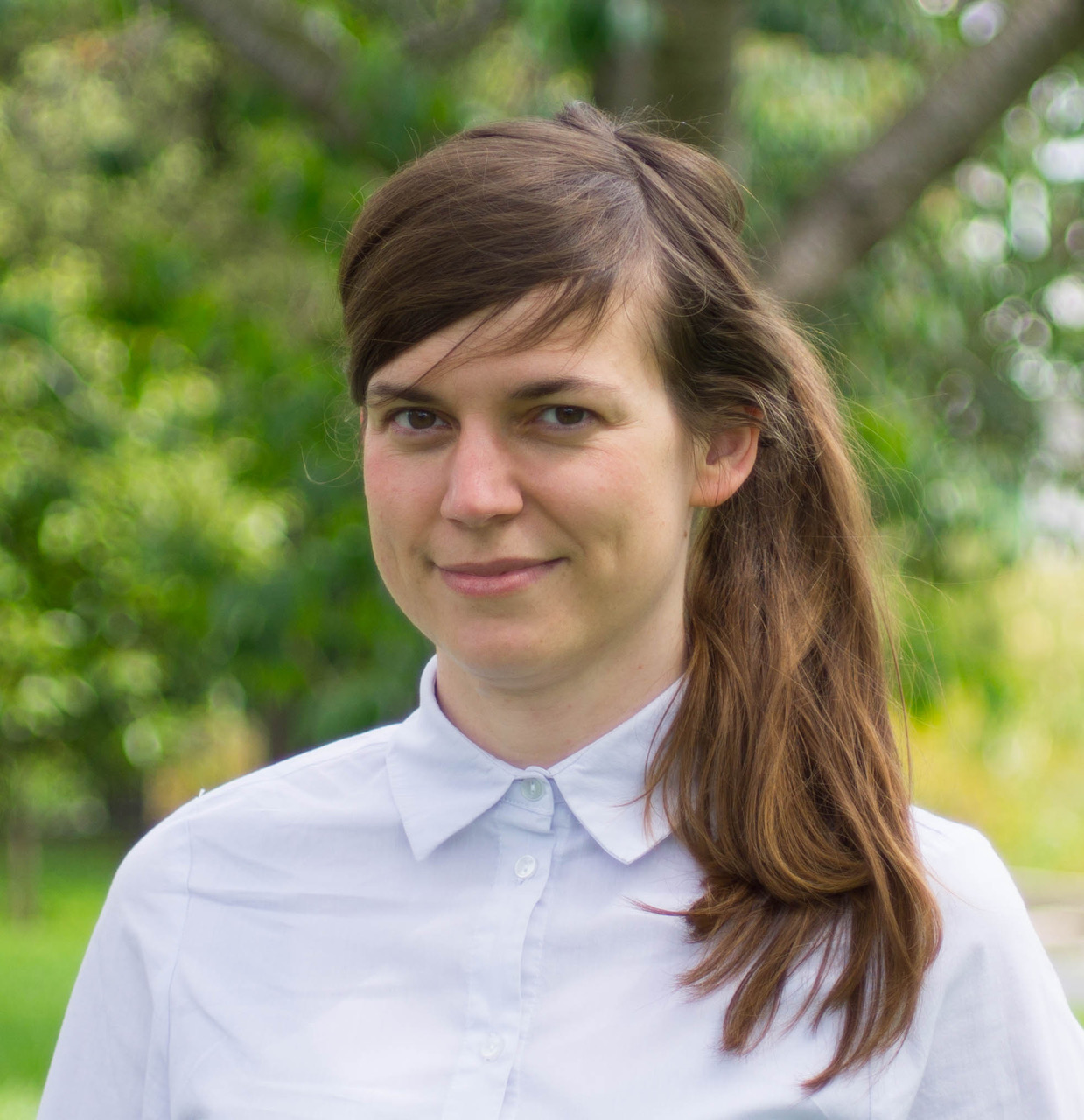}}]{Stefanie Zollmann}
is an Associate Professor at the University of Otago in New Zealand. Before, she worked at Animation Research Ltd on XR visualization and tracking technology for sports broadcasting. She worked as postdoctoral researcher at the Graz University of Technology where she also obtained a PhD degree in 2013. Her main research interests are XR for sports and media, visualization techniques for augmented reality, but also include capturing for XR. 
\end{IEEEbiography}

\begin{IEEEbiography}[{\includegraphics[width=1in,height=1.25in,clip,keepaspectratio]{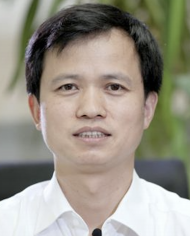}}]{Shi-Min Hu} received the Ph.D. degree from Zhejiang University in 1996. He is currently a Professor with the Department of Computer Science and Technology, Tsinghua University, Beijing. He has published over 100 articles in journals and refereed conference. His research interests include digital geometry processing, video processing, rendering, computer animation, and computer-aided geometric design. He is on the editorial board of several journals, including the IEEE Transactions on Visualization and Computer Graphics, Computer Aided Design, and Computer and Graphics. He is the Editor-in-Chief of Computational Visual Media.
\end{IEEEbiography}





\end{document}